\documentclass[useAMS,usenatbib,usegraphicx]{mn2e}

\usepackage{amssymb}

\title[Infall Signatures in G31 HMC]{Signatures of infall motions in the images of the molecular emission of G31 HMC}

\author[Mayen-Gijon et al. 2013]{J.M. Mayen-Gijon,$^1$ G. Anglada,$^1$ M. 
Osorio,$^1$ L.F. Rodr\'\i guez,$^2$ S. Lizano,$^2$
\newauthor 
J.F. G\'omez,$^1$ and C. Carrasco-Gonz\'alez$^{2,3}$\\
$^{1}$Instituto de Astrof\'{\i}sica de Andaluc\'{\i}a, CSIC, 
Glorieta de la Astronom\'{\i}a s/n, E-18008 Granada, Spain\\
$^{2}$Centro de Radioastronom\'{\i}a y Astrof\'{\i}sica, 
Universidad Nacional Aut\'onoma de M\'exico,
Apartado Postal 72-3 (Xangari), \\ 58089 Morelia, Michoac\'an, Mexico\\
$^{3}$Max-Planck-Institut f\"ur Radioastronomie (MPIfR), Auf dem H\"ugel 69, 53121 Bonn, Germany \\
}

\begin{document}

\date{Accepted. Received}

\pagerange{\pageref{firstpage}--\pageref{lastpage}} \pubyear{ }

\maketitle

\label{firstpage}

\begin{abstract} Although gravitational collapse is supposed to play an 
essential role in the star formation process, infall motions have been 
always elusive to detect. So far, only a few observational signatures 
have been commonly used to claim for the presence of infall. Often these 
features consist in either ``blue-asymmetries'' or absorption at 
red-shifted velocities (e.g., inverse P-Cygni profiles). Both signatures 
are based only on the shape of the line profile and they do not 
guarantee by themselves the presence of dominant infall motions. More 
robust ``mapping signatures'' can be obtained from images that angularly 
resolve the infalling gas. Here we present VLA observations of the 
ammonia inversion transitions (2,2), (3,3), (4,4), (5,5), and (6,6) 
towards the hot molecular core (HMC) near G31.41+0.31 that show the 
signatures of protostellar infall theoretically predicted by 
\citet{ang91}. The intensity of the ammonia emission is compact and 
sharply increases towards the centre in the blue-shifted velocity 
channel maps, while it shows a more flattened distribution in the 
red-shifted velocity channels. Additionally, the emission becomes 
more compact with increasing (relative) velocity for both red and 
blue-shifted channels. We introduce a new infall signature, the 
``central blue spot'', easily identifiable in the first-order moment 
maps. We show that rotation produces an additional, independent 
signature, making the distribution of the emission in the channel maps 
asymmetric with respect to the central position, but without masking the 
infall signatures. All these mapping signatures, which are identified 
here for the first time, are present in the observed ammonia transitions 
of G31 HMC.
 \end{abstract}


\section{Introduction}

It is believed that the star-formation process should be dominated by 
gravitational infall motions that accrete material from the ambient 
cloud into a central protostellar object. Establishing the nature of 
infall motions and distinguishing them from other systematic motions in 
the cloud are not easy tasks, and they have been the subject of many 
papers in the last decades. However, it is not only important to detect 
signatures of infall motions, but also to obtain spatially-resolved 
information that allows us to further investigate the kinematics and 
physical parameters of the molecular core around the protostar. 
Characterizing the infall signatures acquires special interest in the 
case of the formation of massive stars, where different scenarios have 
been proposed (merging of less massive stars, competitive accretion, or 
monolithic collapse; e.g., \citealt{zin07}) and, therefore, where 
obtaining spatially-resolved kinematic information can help to 
discriminate between these scenarios.

\citet{sne77} proposed as evidence of collapse the asymmetric 
self-reversed CO line profiles seen towards some embedded stars in dense 
interstellar clouds. Their modelling of a contracting cloud predicts a 
double-peaked CO line profile where the dip between the two peaks is 
red-shifted with respect to the systemic velocity of the cloud. Soon 
afterwards, this result was criticized by \citet{leu77}, who argued that the 
Sobolev approximation used in the \citet{sne77} calculations was invalid 
for the velocity fields considered, and concluded that the signature was 
not unequivocal for collapse.

\citet{ang87} studied the infall signatures of an angularly unresolved 
spherically symmetric protostellar core under the assumptions of 
optically thick line emission and gravitational infall motions 
dominating the kinematics over turbulent and thermal motions. 
Assuming that in the inside-out gravitational collapse towards a central 
object the velocity increases as the radius decreases, the points with 
the same line-of-sight (l.o.s.) velocity will form closed 
surfaces (isovelocity surfaces) \citep[][see Fig. 1a]{kui78, 
ang87}. All the isovelocity surfaces have the same shape and decrease in 
size with increasing magnitude of the l.o.s. velocity. The 
observed flux density as a function of l.o.s. velocity (i.e., the 
line profile) is determined by the integrated intensity of the 
corresponding isovelocity surfaces. In general, a given line-of-sight 
intersects the same isovelocity surface twice. If the opacity is high 
enough, the observed emission is dominated by the emission coming from 
the side of the isovelocity surface facing the observer (thick line in 
Fig. 1a), while the emission from the rear side (thin line in Fig. 1a) 
remains hidden. Since in protostellar collapse the temperature increases 
towards the centre of the core, it turns out that the blue-shifted 
emission comes from points that, on the average, are closest to the 
protostar (and, therefore, hotter) than the corresponding red-shifted 
ones (see Fig. 1a).
 This results 
 in asymmetric line profiles, 
with the blue-shifted side stronger than the red-shifted one. The 
assumption of a large opacity gradient allows to adopt a simple approach 
similar to the Sobolev approximation (see Appendix A.2 in 
\citealt{ang87}), as only a very thin edge of the isovelocity surface, 
where physical properties are unlikely to vary significantly, would be 
observable. Optically thin lines would have symmetric profiles.

\begin{figure*}
\includegraphics[width=\linewidth]{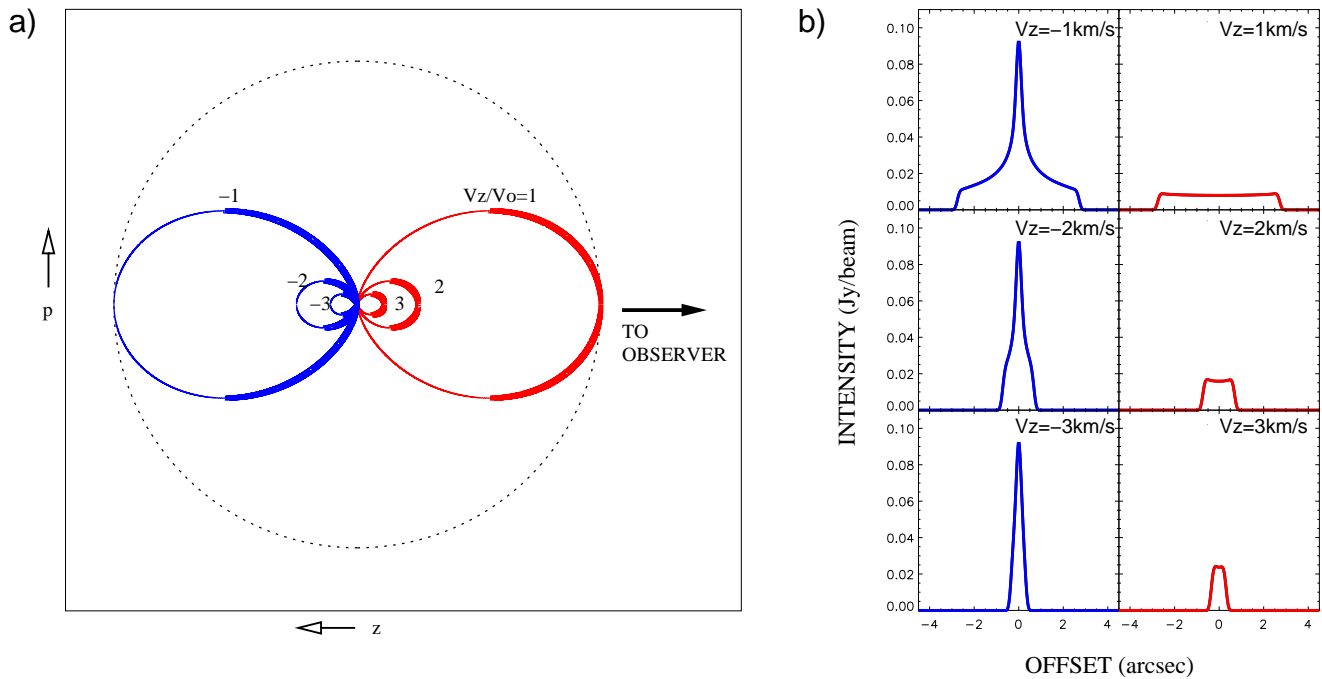}
 \caption{{\em (a)} Surfaces of equal line-of-sight velocity for a 
collapsing protostellar envelope. The (0,0) coordinates correspond 
to the position of the central protostar. Note that a given 
line-of-sight intersects an isovelocity surface at two points. If the 
emission at each velocity is optically thick, only the part of the 
surface facing the observer (thick line) is observable, while the rear 
part (thin line) is not observable. {\em (b)} Intensity as a function of 
angular offset from the source centre for channel maps of different 
velocities. Channel maps are, in fact, images of the temperature 
distribution in the corresponding isovelocity surfaces. For red-shifted 
channels, the centre of the image corresponds to a region slightly 
farther away from the centre of the core (and, thus, slightly colder) 
than the edges. Conversely, for blue-shifted channels, the centre of the 
image arises from a region very close to the centre of the core (and, 
therefore, very hot). This produces very different images in red-shifted 
and blue-shifted channels. In a red-shifted channel, the intensity is 
almost constant over the emitting region (decreasing slightly at the 
centre) while in the corresponding blue-shifted channel the intensity 
increases sharply towards the centre of the image. Additionally, the 
emission becomes more compact as the absolute value of the velocity 
increases (from \citealt{ang91}). Calculations correspond to a simple 
model for an 1 M$_\odot$ protostar at 160 pc observed with a beam 
of $0\farcs2$ at the time when a central mass of $\sim0.5$ 
M$_\odot$ has been accreted.}
 \label{fig1}
 \end{figure*}

\citet{zho93} reported observations of molecular lines towards B335 
showing asymmetric line profiles with the blue side stronger than the 
red side. This result was considered as evidence of gravitational 
protostellar infall since the strengths and profiles of the observed 
lines were in agreement with the predictions of the inside-out collapse 
model of \citet{shu77}. Since then, numerous surveys have been carried 
out and ``blue asymmetries'' in line profiles have been 
extensively interpreted as infall signatures (e.g., \citealt{gre97}; 
\citealt{mar97}; \citealt{lee01}; \citealt{full05}; \citealt{chen10}), 
although in most cases a detailed comparison with the predictions of a 
self-consistent collapse model has not been performed. In several cases, 
inverse P-Cygni profiles have been observed in molecular lines against a 
background bright HII region (e.g., \citealt{ket87,ket88}; 
\citealt{ho96}; \citealt{zha97}; \citealt{zha98}) or against the bright 
dust continuum emission of the hot protostellar core itself (e.g., 
\citealt{dif01}; \citealt{gir09}). These red-shifted absorption features 
are indicative of red-shifted motions of foreground gas and, 
along with additional evidence, have been interpreted as infall 
motions of the surrounding envelope.

However, the ``spectral signatures'' of infall are mostly based on 
the line profile of the overall observed emission, and do not provide 
enough information on the spatial structure of the source. So, other 
scenarios such as rotation, outflow, intrinsic asymmetries in the 
source, or the relative motion of unbound clouds are capable of causing 
similar features. A more complete information can be obtained from 
``mapping signatures'' in images that angularly resolve the source. The 
use of centroid velocity maps, that give some spatial information, has 
been proposed to identify rotation signatures (e.g., the so-called 
``blue bulge'' signature) in infalling cores (e.g., \citealt{ade88}; 
\citealt{nay98}) but in this case the shape of the line profile is not 
taken into account, resulting in an incomplete information on the 
kinematic and physical structure of the core.

A more simple and complete mapping signature of infall can be obtained 
from images that angularly resolve the emission as a function of 
velocity. \citet{ang91} (hereafter A91) investigated the signatures of 
infall in the images obtained from molecular line observations. These 
authors calculated the predicted intensity maps for different 
l.o.s. velocities (channel maps) following the same assumptions 
considered in \citet{ang87}. The intensity map at a given l.o.s. 
velocity results from the emission of the corresponding isovelocity 
surface and, if the opacity is high enough, it is in fact an image of 
the excitation temperature distribution in the side of this isovelocity 
surface facing the observer. For red-shifted channels, the centre of the 
image corresponds to a region that is slightly farther away from the 
centre of the core (and, thus, colder) than the edges (Fig. 1a). 
Conversely, for blue-shifted channels, the centre of the image 
corresponds to the emission from a region very close to the centre of 
the core (and, therefore, very hot). This produces very different images 
in red-shifted and blue-shifted channels. In the image of a red-shifted 
channel, the intensity is almost constant over the emitting region, 
decreasing slightly towards the centre, while in the corresponding 
blue-shifted channel the intensity increases sharply towards the centre 
of the image (see Fig. 1b). Additionally, since the size of the 
isovelocity surfaces decreases with increasing (relative) 
velocity, the emission becomes more compact at higher (relative) 
velocities for both, red- and blue-shifted channels. The 
detectability of these infall signatures depends strongly on the beam to 
source size ratio since one needs to resolve angularly the infalling gas 
for this purpose.
 
Here we present evidence of the signatures proposed by A91, as well 
as additional infall and rotation mapping signatures in the hot 
molecular core (HMC) close to the ultra-compact HII (UCHII) region 
G31.41+0.31 (hereafter G31 HMC). HMCs are small, dense, hot molecular 
clumps usually found in the proximity of UCHII regions, and are believed 
to trace one of the earliest observable stages in the life of massive 
stars. G31 HMC, located at a distance of 7.9 kpc (\citealt{cesa98}), is 
one of the hottest molecular cores discovered so far, and it probably 
harbours an O type protostar. It exhibits strong ammonia emission of 
high excitation transitions (\citealt{chur90}; \citealt{cesa92,cesa98}), 
making G31 HMC one of the few sources where a high signal-to-noise ratio 
analysis of the spatially resolved molecular emission can be attempted. 
\citet{cesa10} reported a double radio continuum source towards the 
centre of the ammonia core, that could trace the two jet lobes from a 
single central protostar or a binary system. Infall motions in G31 HMC 
have already been inferred by \citet{gir09} through the detection of an 
inverse P-Cygni profile against the bright dust emission peak.  
\citet{oso09} carried out a detailed modelling of the source. This 
model, which consists of a spherical envelope collapsing onto an O type 
star, is able to reproduce the observed spectral energy distribution 
(SED) of the source as well as the spectra of the ammonia (4,4) 
transition obtained in the subarcsecond angular resolution VLA 
observations carried out by \citet{cesa98}. In this paper we present 
new, high angular resolution VLA observations of the (2,2) to (6,6) 
ammonia transitions that reveal clear differences between the red- and 
blue-shifted channel maps. In Section 2 we describe our observations. 
In Section 3 we compare the observational results with those of 
the simple model of protostellar infall by A91 and with those of the 
more detailed modelling of G31 HMC by \citet{oso09}; we introduce 
an additional infall signature (the ``central blue spot''), easily 
identifiable in the first-order moment maps; and we study the rotation 
signatures in the spatial intensity profiles, and their presence in G31 
HMC. In Section 4 we give our conclusions.

\section{Observations}

Observations of the ammonia (2,2), (3,3), (5,5) and (6,6) inversion 
transitions were carried out on 2009, March 1 and 3, with the VLA of 
the NRAO\footnote{The National Radio Astronomy Observatory is a 
facility of the National Science Foundation operated under cooperative 
agreement by Associated Universities, Inc.} in the B-configuration 
(VLA Project Code: AM981) providing an angular resolution of 
$\sim0\farcs3$. The phase centre was set to RA(J2000) = $18^{h} 47^{m} 
34.506^{s}$; DEC(J2000) = $-01\degr 12\arcmin 42.97\arcsec$. We used 
the 4IF correlator mode with a bandwidth of 6.25 MHz ($\sim$80 km 
s$^{-1}$) and 31 channels of 195 kHz ($\sim$2.4 km s$^{-1}$) width 
each, plus a continuum channel that contains the central 75 per cent of the 
total bandwidth. This configuration allowed us to observe two 
inversion transitions simultaneously. To make sure that at least the 
main line and one pair of satellites fall within the observational 
bandwidth we centred the observation at a velocity in between the 
main line and the inner red-shifted satellite line.

 Phase, flux and bandpass calibrators were 1851+005, 3C286, and 1773-130 
respectively. Pointing corrections were derived from X-band observations 
for all the calibrators and applied on-line. At the time of the 
observations, the VLA interferometer was going through a transition period 
to turn into the EVLA and 20 out of 27 antennas at that time were already 
EVLA antennas while the rest remained VLA antennas. The frequencies for 
the NH$_{3}$(5,5) and NH$_{3}$(6,6) inversion lines fell very close to the 
edge of the bandwidth of the VLA receivers and only the EVLA antennas were 
used for these two transitions. For the NH$_{3}$(2,2) and NH$_{3}$(3,3) 
inversion lines both EVLA and VLA antennas were used.

Due to the narrowness of the bandwidth used, and because lines in this 
source are extremely broad, line-free channels were not enough to subtract 
the continuum properly and additional continuum observations (VLA Project 
Code: AM994) were carried out on 2009, May 16 for this purpose. In this 
continuum observation we used the same flux and phase calibrators as in 
the line observations.

We also analysed VLA archival data of the NH$_{3}$(4,4) transition taken 
in the A configuration (VLA Project Code: AC748; \citealt{cesa10}). These 
observations were carried out on 2004, October 15, 23, and November 5, 
using the Fast Switching technique with a 80/40 sec source-calibrator 
cycle. The bandwidth was 12.5 MHz ($\sim$160 km s$^{-1}$), with a channel 
width of 195 kHz ($\sim$2.4 km s$^{-1}$).

For all the transitions, data editing and calibration were carried out 
using the Astronomical Image Processing System (AIPS) package of NRAO 
following the standard high-frequency VLA procedures. We also followed 
the advice provided by the NRAO for observations with mixed EVLA and VLA 
antennas. Continuum was subtracted using the task UVSUB of AIPS. This 
correction was important only towards the position of the UCHII region 
and was negligible towards the HMC. Maps were obtained with natural 
weighting and were restored with a circular beam. For the (4,4) 
transition, that was observed in the A configuration, maps were obtained 
using a $uv$ tapering of 1300 k$\lambda$ to improve the signal-to-noise 
ratio. The observational set-up for all the line observations is 
summarized in Table~1.

\begin{table*}
\begin{center}
\caption{Observational parameters of the ammonia observations\label{tab1}}
\begin{tabular}[\textwidth]{cccccccc} 
\\ \hline \hline
&
&
&{Rest}
&
&{Spectral}
&{Synthesized}
&\\
{Transition}
&{VLA}
&{Antenna$^{a}$}
&{Frequency}
&{Bandwidth}
&{Resolution}
&{Beam$^{b}$}
&{rms}\\
{($J$,$K$)}
&{Configuration}
&{Type}
&{(GHz)}
&{(MHz)}
&{(km~s$^{-1}$)}
&{($''$)}
&{(mJy~beam$^{-1}$)} \\
\hline
(2,2) & B & VLA(7)/EVLA(20) &23.7226333  &6.25 & 2.468 &0.33                    &0.7 \\
(3,3) & B &VLA(7)/EVLA(20)  &23.8701292  &6.25 & 2.453 &0.33                    &0.7 \\
(4,4) & A &VLA(27)          &24.1394163  &12.5 & 2.426 &0.16$^{c}$ &0.5 \\
(5,5) & B &EVLA(20)         &24.5329887  &6.25 & 2.386 &0.37                    &0.9 \\
(6,6) & B &EVLA(20)         &25.0560250  &6.25 & 2.337 &0.34                    &1.0 \\
\hline
\end{tabular}
\end{center}
$^{a}${Numbers in parenthesis indicate the number of antennas of either type.} \\
$^{b}${HPBW of the circular restoring beam.} \\
$^{c}${Interferometric visibilities have been tapered to increase the signal-to-noise ratio.}
\end{table*}

In Fig. 2 we show an overlay of the maps of the zero-order moment 
(integrated intensity) and first-order moment (intensity weighted 
mean velocity) of the observed ammonia (2,2), (3,3), (4,4), (5,5), 
and (6,6) main line emission.

\begin{figure*}
\includegraphics[height=0.93\textheight]{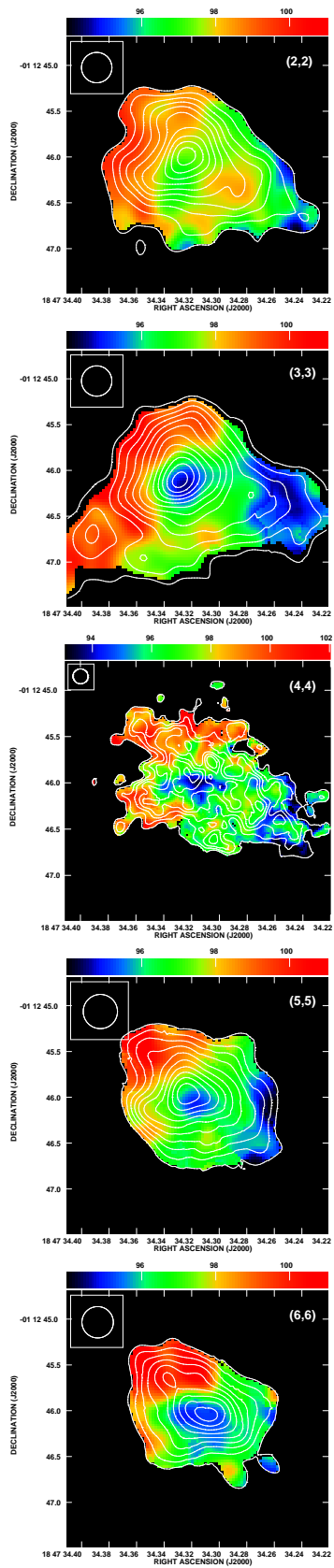}
 \caption{Overlay of the integrated intensity (zero-order moment; 
contours) and the intensity weighted mean velocity (first-order moment; 
color scale) maps of the ammonia (2,2), (3,3), (4,4), (5,5), and (6,6) 
main line towards G31 HMC. Contour levels are 25, 30, 45, 60, 75, 90, 
105, 120, 135, 150, 165, 180, 195, 210 mJy beam$^{-1}$ km s$^{-1}$. The 
synthesized beam is shown in the upper left corner of each map. Color 
scale ranges from 93.2 to 102 km s$^{-1}$. Note the spot of blue-shifted 
emission towards the integrated intensity peak in all the maps.
 }
 \label{fig2}
 \end{figure*}

\section{Results and discussion}

We detected ammonia emission in all the five observed transitions. This 
is the first interferometric detection of the ammonia (3,3), (5,5), and 
(6,6) transitions in this source. In the maps (Fig. 2), the 
ammonia emission appears as a compact condensation, clearly associated 
with the double radio continuum source (angular separation = $0\farcs2$) 
reported by \citet{cesa10}. The peak brightness temperatures are 
$\sim$100-150 K for all the observed ammonia transitions, similar to the 
values obtained previously for the (4,4) transition 
\citep{cesa98,cesa10}. The emission of the main line is optically thick 
in the five transitions studied, with satellite to main line intensity 
ratios close to unity. Lines are extremely broad ($\sim$ 13 km 
s$^{-1}$), so that the inner and outer satellite lines blend together 
even for the (2,2) transition, where the separation in velocity between 
satellite lines is $\sim$ 9 km s$^{-1}$. In this paper we present an 
analysis of the intensity as a function of the projected distance to the 
core centre (which we call as ``spatial intensity profile''), in order 
to identify possible differences between red-shifted and blue-shifted 
channels. A more extensive analysis of the observations will be 
presented in a future paper (Mayen-Gijon et al. in preparation).

\subsection{Spatially resolved infall signatures}

\subsubsection{Spatial intensity profiles}

To search for the spatially resolved infall signatures (see section 1) 
we need to compare the observed intensity as a function of distance to 
the centre of the core for pairs of velocity channels symmetrically 
located at both sides of the systemic velocity. In order to improve the 
signal-to-noise ratio, for every channel map we averaged the emission of 
the main line over circular annuli of different radii and 0$\farcs$15 
width, and then we plotted the average intensity as a function of the 
radius. The centre was chosen to be at RA(J2000) = $18^{h} 47^{m} 
34.311^{s}$; DEC(J2000) = $-01\degr 12\arcmin 45.90\arcsec$, which 
corresponds to the position of the peak of the integrated emission of 
the ammonia (6,6) satellite lines. We have chosen the position of the 
satellite lines of the (6,6) transition because they have the lowest 
opacity and highest excitation among the observed lines, so they can 
penetrate further inside the core, and may trace better the warmer gas, 
close to the central heating object. Our chosen central position lies 
roughly in between the positions of the two embedded radio 
continuum sources detected by \citet{cesa10}, which are separated $\sim 
0\farcs2$.

The systemic velocity of the cloud is not accurately determined. In the 
literature we find velocities ranging from 96.24 km s$^{-1}$ 
(\citealt{bel05}; from ground state CH$_3$CN (12$-$11) observations) to 
98.8 km s$^{-1}$ (\citealt{cesa94}, from low resolution NH$_3$(4,4) 
observations). We have adopted the value of V$_{LSR}$=97.4 km s$^{-1}$ 
from the interferometric data of \citet{bel05}, obtained by fitting 
simultaneously several K components of the $v_8=1$ CH$_3$CN (5-4) 
transition.

As the original purpose of the observations presented here was not to 
search for infall signatures, they were not designed adequately for this 
study. The spectral resolution is poor ($\sim$ 2.4 km s$^{-1}$) and the 
observations of the different transitions were not centred exactly at 
the same velocity. Thus, we cannot obtain pairs of velocity channels 
that are exactly symmetric in velocity with respect to the rest velocity 
of the cloud, which hinders an accurate comparison of red- and 
blue-shifted channel pairs. As our best approximation, we have compared 
channel pairs with the most similar velocity shifts with respect to the 
assumed systemic velocity of the cloud. Given the uncertainty in this 
central velocity and the large channel widths ($\sim2.4$ km s$^{-1}$), 
in order to ensure that we are comparing emission with different 
velocity signs (i.e., red vs. blue) we have excluded the channel pair 
closest ($\sim \pm 2$ km s$^{-1}$) to the systemic velocity.

Fig. 3 shows the observed intensity as a function of the projected 
distance to the core centre for blue-shifted (blue continuous lines) and 
red-shifted (red dotted lines) velocity channels of the ammonia (2,2), 
(3,3), (4,4), (5,5) and (6,6) main lines. The main lines of all these 
transitions are certainly optically thick (as indicated by the main to 
satellite line ratio that is close to unity), and a pattern is clearly 
identifiable in the images of all the observed transitions. For each 
pair of blue/red channels, the intensity of the blue-shifted channel is 
stronger, and increases more sharply towards the centre, than in the 
corresponding red-shifted channel, where the intensity distribution is 
flatter. Moreover, as the velocity increases with respect to the 
systemic velocity, the emission becomes more compact (a narrower spatial 
intensity profile). This behaviour is in agreement with the expected 
mapping signatures of infall for angularly resolved sources described by 
A91 (see above, and Fig. 1), suggesting that gravitational infall 
motions strongly contribute to the kinematics of the G31 HMC. Since the 
satellite lines appear to be also optically thick (at least for the 
lowest excitation transitions), similar signatures would be expected in 
the satellite line channel maps. Unfortunately, because of blending 
between satellite lines due to the large line widths, a similar analysis 
cannot be carried out for the satellite emission.

\begin{figure*}
\includegraphics[width=\linewidth]{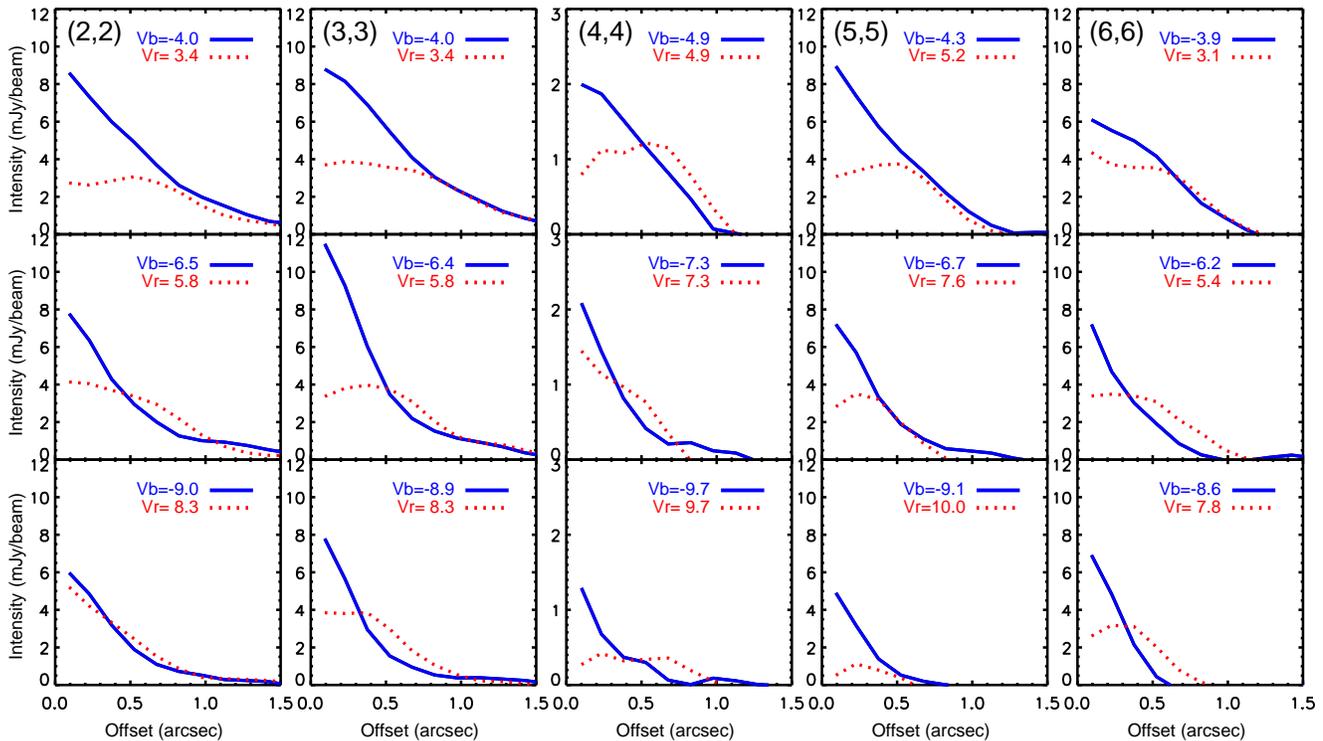}
\caption{Circularly-averaged observed intensity as a function of angular 
distance to the centre of G31.41 HMC, for different pairs of 
blue-shifted (solid blue line) and red-shifted (dotted red line) channel 
maps. Results for the ammonia (2,2), (3,3), (4,4), (5,5) and (6,6) 
transitions (from left to right) are shown. The sizes of the synthesized 
beams are given in Table 1. Labels indicate the channel velocity (in 
km~s$^{-1}$) relative to the assumed systemic velocity of the cloud 
($V_{\rm LSR}=97.4$ km~s$^{-1}$).}
 \label{fig3}
 \end{figure*}

To confirm the presence of these infall signatures more quantitatively, 
we compare our observational results with the predictions of the model 
for G31 HMC developed by \citet{oso09}. These authors modelled G31 
HMC as a spherically symmetric envelope of dust and gas that is 
collapsing onto a recently formed massive star ($M_*\simeq$ 25 
M$_\odot$) that is undergoing an intense accretion phase ($\dot M\simeq 
3\times10^{-3}$ M$_\odot$ yr$^{-1}$). The envelope consists of an inner, 
infalling region ($r_{\rm ew}\simeq 2.3\times10^4$ au) that is 
surrounded by a static part ($R_{\rm ext}\simeq 3\times10^4$ au). The 
density and velocity structure of the envelope are obtained from the 
solution of the dynamical collapse of a singular logatropic sphere. The 
stellar radiation ($L_* \simeq 8\times10^4$ L$_\odot$) and the accretion 
luminosity ($L_{\rm acc} \simeq 1.5\times10^5$ L$_\odot$) provide the 
source of heating of the envelope, whose temperature as a function of 
the distance to the centre is self-consistently calculated from the 
total luminosity. The excitation and absorption coefficient of the 
ammonia transitions are calculated according to the physical 
conditions along the envelope, and the radiative transfer is performed 
in order to obtain the emerging ammonia spectra. The parameters of the 
model are determined by fitting the observed spectral energy 
distribution (SED) and the VLA B-configuration NH$_3$(4,4) line spectra 
obtained in the observations of \citet{cesa98} that angularly resolve 
the source. Fig. 4 shows the intensity predicted by the G31 HMC 
model as a function of the projected distance to the core centre, for 
different pairs of blue/red-shifted channels of the ammonia (2,2) to 
(6,6) main lines. The angular and the spectral resolutions have been set 
equal to those of the observations.

\begin{figure*}
\includegraphics[width=\linewidth]{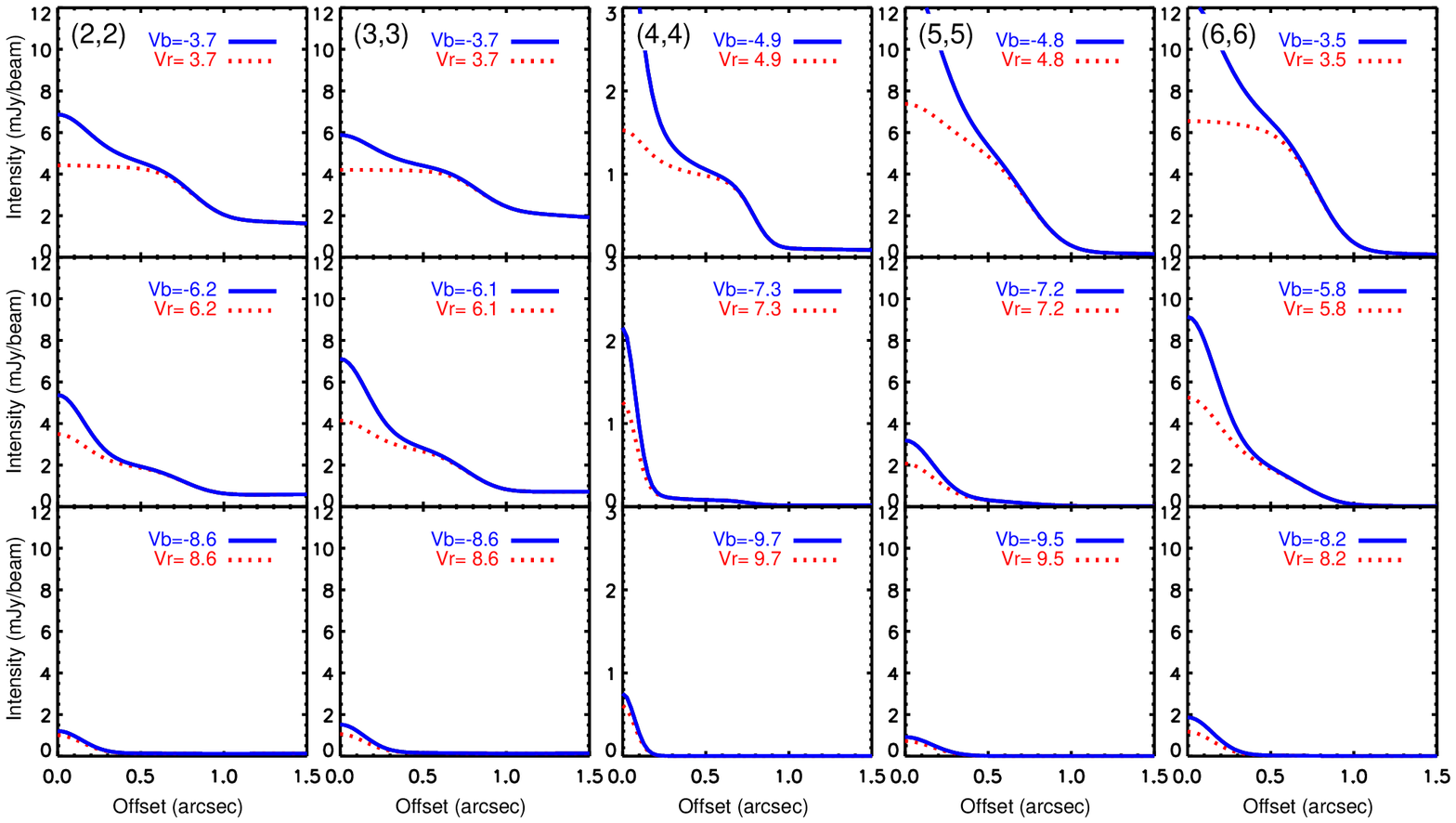}
 \caption{Intensity as a function of angular distance to the centre, 
obtained with the model of G31 HMC by \citet{oso09}, for different pairs 
of blue-shifted (solid blue line) and red-shifted (dotted red line) 
channel maps. Results for the ammonia (2,2), (3,3), (4,4), (5,5) and 
(6,6) transitions (from left to right) are shown. Model results have 
been convolved with a Gaussian with a FWHM equal to that of the 
observational beam (see Table 1). Labels indicate the velocity (in 
km~s$^{-1}$) relative to the systemic velocity of the cloud.}
 \label{fig4}
\end{figure*}

As can be seen from Fig. 4, the spatial intensity profiles 
predicted by the model of G31 HMC developed by \citet{oso09} show 
the basic signatures of spatially resolved infall proposed by A91 from a 
simplified, general treatment of protostellar infall. The spatial 
intensity profiles of the blue-shifted channels are centrally peaked, 
while those of the red-shifted channels are flatter; and the emission in 
both red- and blue-shifted channels becomes more compact as the relative 
velocity increases. Furthermore, by comparing Figs. 3 and 4 
we note that there is a remarkable quantitative agreement between the 
observations of G31 HMC and the model of this source, as the shape 
and range of values are similar in the observed and modelled spatial 
intensity profiles. Thus, we conclude that the general infall 
signatures described by A91 are present in G31 HMC, suggesting that 
infall motions have a significant contribution to the kinematics of this 
source. The quantitative agreement with the predictions of a specific 
model for G31 HMC indicates that these signatures are sensitive to the 
physical properties (in particular, the kinematics and temperature 
profile) of the source and, therefore, that in principle they can be 
used to verify and/or improve source models.

Although there is a general agreement, the precise shape of the 
predicted and observed intensity profiles are not fully coincident. For 
example, at high velocities (relative to rest) the model emission 
is in general weaker and more compact than observed (bottom panels 
in Figs. 3 and 4). Some of the discrepancies could be due to the 
fact that the pairs of velocity channels extracted from the observations 
are not exactly symmetric in velocity because of the large channel 
widths and the uncertainty in the central velocity. Observations with 
higher spectral resolution of optically thick lines, combined with 
additional observations of optically thin lines (to accurately determine 
the central velocity) are required to better study the trends presented 
here. Presence of rotation and outflow motions (e.g., \citealt{cesa11}) 
and, in general, departures from spherical symmetry, that are not 
considered in the modelling of Osorio et al. (2009) (e.g., the 
presence of a central binary), could affect the infall signatures, 
and account for some of the observed differences. In Section 3.2 we 
will examine the effects of rotation on the spatial intensity profiles.

We want to stress that G31 HMC is an exceptional source, with very 
strong molecular ammonia emission, making it possible to identify for 
the very first time spatially resolved infall signatures in molecular 
emission. The advent of large interferometers with improved sensitivity, 
such as ALMA, should make it possible to extend this kind of studies to 
a large number of sources in the near future.

\subsubsection{The central blue spot}

The infall signatures in the spatial intensity profiles produce an 
additional signature in the first-order moment (intensity weighted mean 
velocity, equivalent to a centroid velocity map). This signature is 
easily identifiable, and consists in a compact spot of blue-shifted 
emission towards the peak of integrated intensity (zero-order moment) of 
the source. This spot of blue-shifted emission is a consequence of the 
differences in intensity and spatial distribution of the emission in the 
red- and blue-shifted channels. As explained in Section 1 and 
illustrated in Fig. 1, the emission near the edges of the emitting 
region has a similar intensity in the pairs of red- and blue-shifted 
velocity channels that are symmetrically located with respect to the 
systemic velocity of the cloud. However, while the intensity 
distribution remains almost flat in the red-shifted channels, it rises 
sharply towards the centre in the blue-shifted channels. As a 
consequence, the intensity-weighted mean velocity towards the central 
region will appear blue-shifted because of the higher weight of the 
strong blue-shifted emission.  Additionally, the integrated intensity 
(zero-order moment) will peak towards the central position. As one moves 
away from the centre the integrated intensity decreases, the blue and 
red-shifted intensities become similar, and the intensity-weighted mean 
velocity approaches the systemic velocity of the cloud. Therefore, the 
first-order moment of an infalling envelope is expected to be 
characterized by a compact spot of blue-shifted emission towards the 
position of the zero-order moment peak. We will designate this infall 
signature the ``central blue spot''.

The ``central blue spot'' will appear overlapped with other possible 
velocity gradients (e.g., due to rotation) that may be present in the 
first-order moment of the source. This is the case of G31 HMC where the 
first-order moment (Fig. 2) shows both a velocity gradient from NE to 
SW, and a spot of blue-shifted emission towards the peak of the 
zero-order moment. The velocity gradient suggests rotation with a PA 
$\simeq 150^\circ$. The clear detection of the ``central blue spot'' 
signature in G31 HMC indicates that infall motions play a fundamental 
role in the gas kinematics of this source. As can be seen in Fig. 2, the 
blue spot signature is present in all the observed ammonia transitions, 
confirming that its detection is a robust result. It is less evident in 
the (2,2) transition, probably because this transition has a larger 
contribution from outer, colder gas.

We note that the opacity is expected to decrease in the central part of 
the blue-shifted emitting region (see Appendices A and C of Anglada et 
al. 1987). This can reduce the degree of asymmetry and partially erase 
the blue spot signature in molecular transitions of moderately high 
optical depth. The observations and modelling suggest that this is not 
the case in G31 HMC.

We also note that the ``central blue spot'' infall signature is related 
to the ``blue bulge'' signature introduced by Walker, Narayanan, \& Boss 
(1994), who realized that in the centroid velocity map of a source with 
a rotation velocity gradient, the blue-shifted contours cross the 
rotational axis into the side that would otherwise correspond to 
red-shifted emission. The ``blue bulge'' signature is interpreted as a 
consequence of the overall asymmetry between blue- and red-shifted 
emission produced by infall motions. The ``central blue spot'' signature 
introduced here traces specifically the sharp increase in the degree of 
asymmetry between the blue- and red-shifted intensities as one moves 
towards the centre, which produces a central compact blue spot.
The ``blue bulge'' signature is also visible in our data. Fig. 2 shows 
that, for all the observed transitions, part of the blue-shifted 
emission (the emission shown in green color in the figure) extends to 
the NE and penetrates into the side that would otherwise correspond to 
red-shifted emission.

\subsection{On the effects of rotation} 

As can be seen in Fig. 2, the first-order moment (intensity weighted 
mean velocity) of the ammonia emission in G31 HMC shows a clear velocity 
gradient in the SW-NE direction, with average velocities preferentially 
blue-shifted in the SW side and preferentially red-shifted in the NE 
side of the source. This velocity gradient has been previously reported 
by other authors using several molecular transitions (e.g., Beltr\'an et 
al. 2004; Gibb et al. 2004; Cesaroni et al. 2011), and has been 
attributed either to rotation or to a bipolar outflow. In order to 
properly interpret the G31 HMC results, in the following we will discuss 
which are the expected rotation signatures in the spatial intensity 
profiles, and how they affect the infall signatures discussed in the 
previous sections.

To analyse the case of an infalling rotating envelope, and to compare it 
with the non-rotating case, we have considered a system with properties 
similar to those of the example described in A91 (Fig. 1), but with the 
velocity field given by the TSC (Terebey, Shu, \& Cassen 1984) 
prescription. The TSC formalism provides a generalised solution of the 
problem of the spherical collapse of a singular isothermal sphere (Shu 
1977) that includes the effects of an initially uniform and slow 
rotation. In the TSC velocity field (equations 88-90 of Terebey, Shu, \& 
Cassen 1984), not only radial but also azimuthal and polar velocity 
components are present. As a consequence, matter does not fall radially 
onto the central protostar, but settles into a centrifugally supported 
disk around it (Terebey, Shu, \& Cassen 1984; Shu, Adams, \& Lizano 
1987). In the inner region of the infalling envelope (near the 
centrifugal radius), the magnitude of rotation and infall velocities 
tend to equal, and the density adopts a flattened configuration towards 
the equatorial plane. At larger radii (much larger than the centrifugal 
radius), the density law tends to the free-fall form solution with 
nearly radial streamlines.

Fig. 5a shows the contours of equal l.o.s. velocity of the TSC 
velocity field close to the equatorial $z$-$p$ plane, for the example 
considered. The TSC isovelocity contours are closed curves with sizes 
similar to those of radial infall shown in Fig. 1a. However, in radial 
infall all the isovelocity contours are aligned and axially symmetric 
with respect to the line-of-sight towards the centre, while the TSC 
isovelocity contours are asymmetric with respect to this line-of-sight, 
and they appear somewhat distorted. The axis going from the centre of 
the core to the vertex (the point more distant from the centre) of the 
TSC isovelocity contours is rotated with respect to the line-of-sight. 
The angle of rotation of this axis increases from the outer (lower) 
isovelocity contours to the inner (higher) isovelocity contours where it 
reaches $\sim45\degr$. This is because in the TSC formalism the rotation 
velocity is small in the outer parts of the infalling region, that 
behave almost as free-fall, while in the proximity of the centrifugal 
radius, in the equatorial plane, it becomes comparable to the radial 
component.

\begin{figure*}
\includegraphics[width=\linewidth]{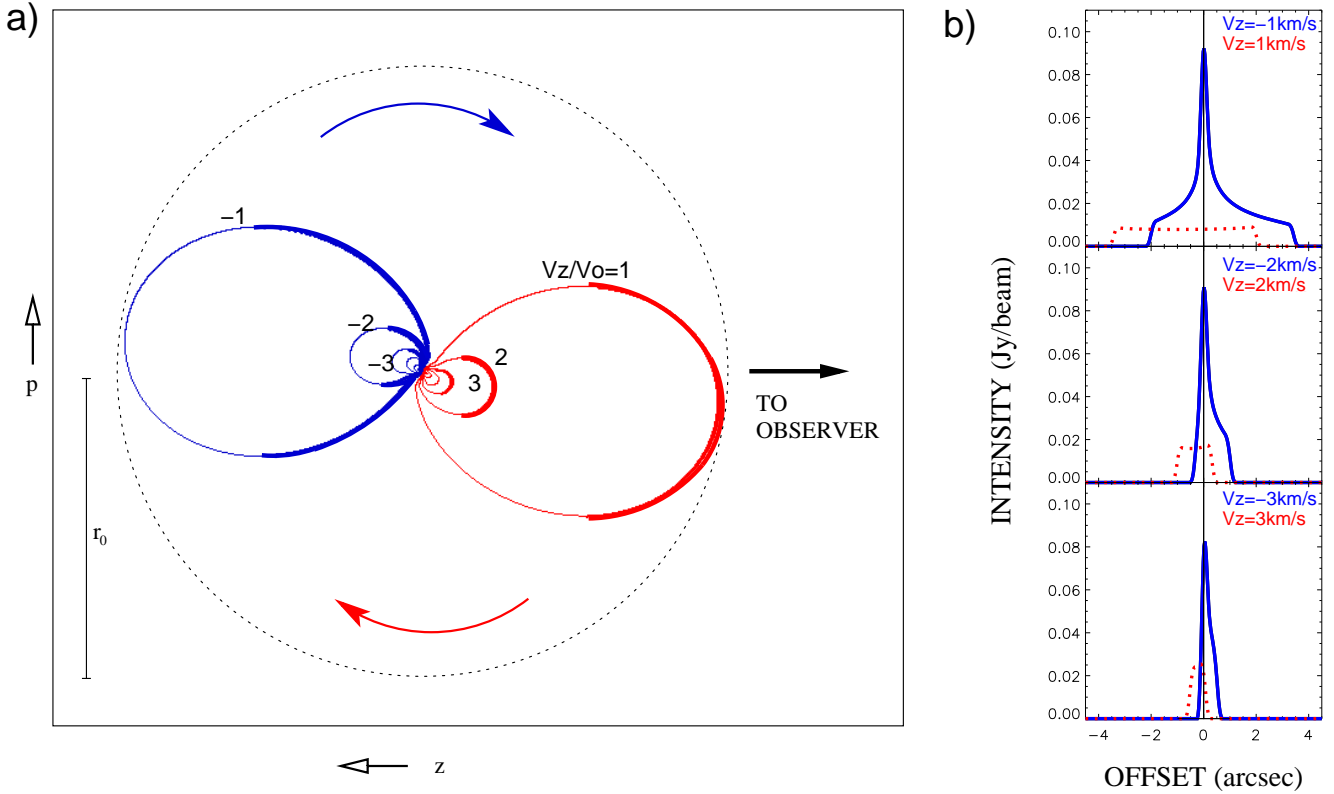}
 \caption{{\em (a)} Contours of equal l.o.s. velocity for a 
collapsing and rotating protostellar envelope in a $z$-$p$ plane close 
to the equator, as described by the TSC formalism 
(Terebey, Shu, \& Cassen 1984). The (0,0) offset corresponds to the 
position of the central protostar. The angular momentum is assumed to be 
perpendicular to the $z$-$p$ plane, and pointing inside the page 
(clockwise rotation, as seen from the reader). Note that a given 
line-of-sight intersects an isovelocity surface at two points. If the 
emission at each velocity is optically thick, only the part of the 
isovelocity surface facing the observer (thick line) is observable, 
while the rear part (thin line) is not observable. In comparison with 
the radial collapse shown in Fig. 1, where all the isovelocity surfaces 
are symmetric with respect to the line-of-sight to the centre, the main 
effect of the rotation of the cloud is to rotate the axes of the 
isovelocity surfaces in the sense of the cloud rotation. {\em (b)} 
Intensity as a function of angular offset from the source centre, in the 
equatorial direction, for channel maps of different l.o.s. velocity. As 
channel maps are, in fact, images of the temperature distribution in the 
corresponding isovelocity surfaces, the misalignment of the axes of the 
isovelocity surfaces with respect to the central line-of-sight results 
in asymmetries in the observed spatial intensity profiles. Calculations 
correspond to a central protostar of $\sim 0.5$ M$_\odot$ at 160 pc 
observed with a beam of $0\farcs2$.  A centrifugal radius of 50 au has 
been assumed.
}
 \label{fig5}
\end{figure*}

Following the simplified analysis of A91, that assumes a high enough 
optical depth, the intensity map at a given l.o.s. velocity is, 
in fact, an image of the excitation temperature distribution in the side 
facing the observer of the corresponding isovelocity surface. Since the 
axes of the TSC isovelocity surfaces are misaligned with respect to the 
line-of-sight, the observer will see an asymmetric source in each 
velocity channel. In the example considered (Fig. 5), the emission in 
the red-shifted channels will be more extended in the eastern side 
(negative offsets) and more compact in the western side (positive 
offsets), while in the blue-shifted channels the emission will appear 
more extended in the western side and more compact in the eastern side. 
 Thus, the spatial intensity profiles will be asymmetric as illustrated 
in Fig. 5b, where we show the spatial intensity profile along the 
equatorial direction, for different velocity channels. As in the 
non-rotating case, the images in blue-shifted channels present centrally 
peaked intensity distributions, while the red-shifted channels present 
flatter intensity distributions; and the emission becomes more compact 
with increasing l.o.s. velocity. Then, the main difference with respect 
to the non-rotating case is the opposite asymmetry, with respect to the 
central position, of the red- and blue-shifted pairs of velocity 
channels.

It can be shown that the asymmetries inferred from the velocity field of 
the TSC collapse are a general property of the rotating infalling 
envelopes. Let us consider a given isovelocity surface in an infalling 
and rotating cloud (in the following descriptions, we consider 
velocities relative to the systemic velocity of the cloud). If rotation 
was absent (pure radial infall, as described in Section 1) the 
isovelocity surface considered, which would correspond to a sole l.o.s. 
infall velocity, would be symmetric with respect to any plane containing 
the centre of the core. However, rotation will produce l.o.s. velocity 
components of opposite sign at both sides of the rotation axis that, 
together with the l.o.s. infall component, will contribute to the total 
l.o.s. velocity. On the side where the l.o.s. rotation and infall 
velocities have the same sign (left side for the red-shifted channels or 
right side for the blue-shifted channels, as seen from the observer in 
Fig. 5a) rotation increases the magnitude of the total l.o.s. velocity. 
Thus, a contribution of the infall component smaller than in the case of 
no rotation would be required to obtain the same total l.o.s. velocity 
considered. Infall isovelocity surfaces are nested like matryoshka 
dolls, with decreasing magnitude of the l.o.s. infall velocity in the 
outer, larger surfaces (Fig. 1a); therefore, this side of the 
isovelocity surface considered will consist of points located at outer 
positions with respect to the case of pure radial infall. On the other 
hand, on the side where l.o.s. rotation and infall velocities have 
opposite sign, a larger contribution of the infall component will be 
required to compensate the decrease in the magnitude of total l.o.s. 
velocity produced by rotation. Therefore, on this side, the isovelocity 
surface considered will consist of points closer to the centre (higher 
magnitude of the l.o.s. infall velocity) than in the case of pure radial 
infall. In summary, the observed emission at a given l.o.s. velocity 
from a collapsing and rotating cloud will be asymmetric with respect to 
the rotation axis, being more extended on the side where rotation and 
infall velocity components have the same sign and more compact on the 
opposite side. Likewise, the spatial intensity profiles are stretched on 
the side where the channel and rotation velocities are both red-shifted 
or both blue-shifted while they are shrunk on the opposite side.

It should also be noted that the set of spatial intensity profiles for 
different velocity channels is actually equivalent to a 
position-velocity (P-V) diagram. However, in a conventional P-V diagram 
the spatially resolved infall signatures are usually more difficult to 
identify.

To study whether the rotation signatures discussed above are in 
qualitative agreement with what is observed in G31 HMC, we obtained the 
average intensity of the ammonia main line as a function of projected 
distance to the centre (ring-averaged values, as in Section 3.1, to 
increase the signal-to-noise ratio), but we calculated this intensity 
separately in the half of the source that rotation shifts to the blue 
(SW) and in the half that rotation shifts to the red (NE). From Fig. 2 
we infer that the rotation axis (i.e., the angular momentum of the core) 
has a PA $\simeq 150\degr$. So, for every observed velocity channel, we 
obtained the average intensity over half-annuli of different radii 
corresponding to the NE ($-30\degr < {\rm PA} < 150\degr$) and SW 
($150\degr < {\rm PA} < 330\degr$) halves of the source. In Fig. 6 we 
plotted the resulting intensity profiles for different velocity channels 
of the observed ammonia transitions. As can be seen in the figure, the 
intensity profiles show the general trends predicted for a rotating 
collapsing core: (i) The spatial intensity profiles of the blue-shifted 
channels sharply increase towards the centre, while those of the 
red-shifted channels are flatter; (ii) The profiles become more narrow 
as the magnitude of the infall velocity increases; (iii) The intensity 
profiles are asymmetric with respect to the central position, being the 
NE part (negative offsets) in the red-shifted channels more extended 
than the SE part (positive offsets), while the opposite is true for the 
blue-shifted channels. This behaviour is in agreement with what is 
expected for an infalling source with clockwise rotation (as seen from 
north).

It should be noted that, so far, we only have constructed a spherically 
symmetric model for G31 HMC (Osorio et al. 2009); therefore, we can only 
evaluate the effects of rotation in a qualitative way. We are working on 
an integral modelling of the source, where the observational data are 
fitted to an infalling rotating envelope, whose results will be 
presented elsewhere (Mayen-Gijon et al., in preparation).

\begin{figure*}
\includegraphics[width=\linewidth]{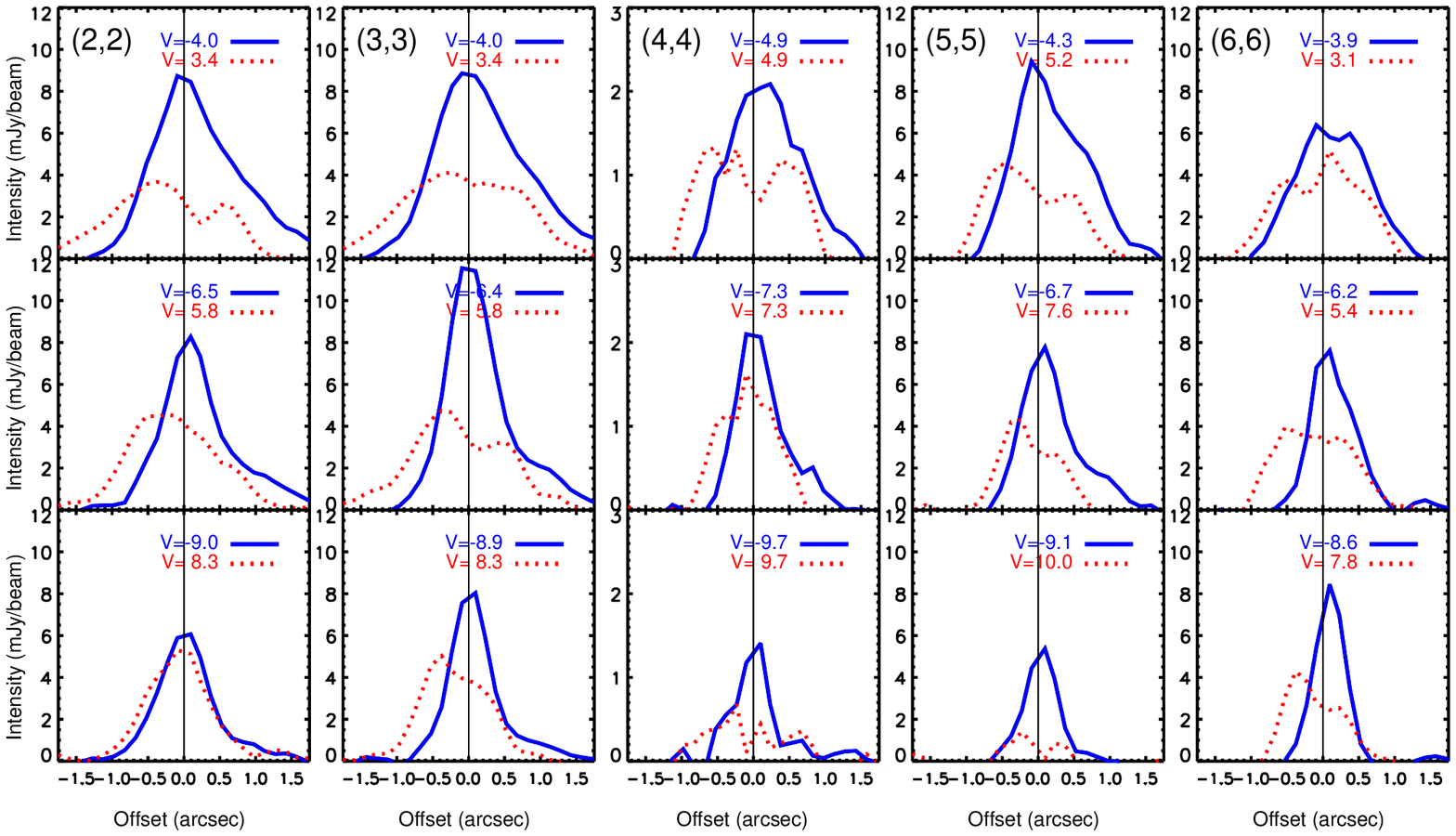}
 \caption{Observed intensity as a function of angular offset from the 
centre of G31.41 HMC, for different pairs of blue-shifted (solid blue 
line) and red-shifted (dotted red line) channel maps. Intensities have 
been averaged over half-annuli at both sides of the rotation axis, 
assumed to be at PA = 150$\degr$. Negative offsets correspond to the NE 
half of the source ($-30\degr < {\rm PA} < 150\degr$), where rotation 
velocities are red-shifted, and positive offsets correspond to the SW 
half of the source ($150\degr < {\rm PA} < 330\degr$), where rotation 
velocities are blue-shifted. 
Results for the ammonia (2,2), (3,3), (4,4), (5,5) and (6,6) transitions 
(from left to right) are shown. The sizes of the synthesized beams are 
given in Table 1. Colour labels indicate the channel velocity (in 
km~s$^{-1}$) relative to the assumed systemic velocity of the cloud 
($V_{\rm LSR}=97.4$ km~s$^{-1}$).}
 \label{fig6}
\end{figure*}

Thus, we conclude that spatially resolved rotation and infall signatures 
can be identified in the images of the molecular ammonia emission of G31 
HMC. It is important to note that rotation does not mask the mapping 
signatures of infall proposed by A91, but just modifies these signatures 
by making the spatial intensity profiles asymmetric.

\section{Conclusions}

We found that the high angular resolution images of the ammonia (2,2), 
(3,3), (4,4), (5,5), and (6,6) inversion transitions in G31 HMC present 
the mapping signatures of gravitational infall predicted by A91: (i) 
Given a pair of red/blue-shifted channels at symmetric velocities 
with respect to the systemic velocity of the cloud, in the blue-shifted 
channels the intensity as a function of projected distance to the core 
centre is stronger and sharply increases towards the centre, 
while the red-shifted channels show a flatter intensity distribution; 
(ii) The emission becomes more compact as the velocity increases 
with respect to the systemic velocity of the cloud.

We introduced a third mapping signature of infall, that we call the 
``central blue spot'', consisting in a spot of blue-shifted emission in 
the first-order moment towards the position of the zero-order moment 
peak. We found that the central blue-spot infall signature is also 
present in all the observed ammonia transitions of G31 HMC. These three 
signatures are robust infall signatures that, to our knowledge, have 
been observationally identified here for the first time. These 
signatures cannot be easily mimicked by other motions, and indicate that 
infall motions play a fundamental role in the gas kinematics of G31 HMC.

There is a good quantitative agreement, both in the shape and range of 
values, between the spatial intensity profiles predicted by the model 
for G31 HMC developed by \citet{oso09} and those observed in the channel 
maps of the ammonia (2,2) to (6,6) inversion transitions. This is 
particularly remarkable, as only the SED and (4,4) transition data were 
used in fitting the model parameters.

We studied how rotation affects the spatial intensity profiles of an 
infalling envelope. We concluded that the rotation signature makes the 
spatial intensity profiles asymmetric with respect to the central 
position but it does not mask the A91 infall signatures. The spatial 
intensity profile of the image in a given velocity channel (red- or 
blue-shifted) is stretched towards the side where rotation has the same 
sign (red- or blue-shifted, respectively), and it is shrunk on the 
opposite side.  These rotation signatures are present in the spatial 
intensity profiles of the images of the ammonia transitions observed in 
G31 HMC.

In summary, G31 HMC shows a quantitative agreement with the A91 infall 
signatures, it presents the ``central blue spot'' infall signature, 
and shows a qualitative agreement with the signatures expected in a 
rotating infalling envelope.

G31 HMC is an exceptional source because of its strong ammonia emission 
that allowed us to identify these infall signatures for the first time. 
However, with the advent of new and improved high angular resolution 
facilities, it will become possible to identify these signatures in more 
sources. This kind of observations, combined with a detailed modelling, 
can be used to determine the kinematic properties of infalling 
protostellar envelopes.

\section*{Acknowledgments}

JMM-G, GA, MO, and JFG acknowledge support from MICINN (Spain) 
AYA2008-06189-C03-01 and AYA2011-30228-C03-01 grants (co-funded with 
FEDER funds). SL acknowledges support from PAPIIT-UNAM IN100412. 
We thank the referee, Eric Keto, for his valuable comments. 
We would like to thank Riccardo Cesaroni for his advise in the reduction 
of the ammonia (4,4) data.

\label{lastpage}

\end{document}